\documentclass[fleqn,10pt]{wlscirep}
\usepackage[T1]{fontenc}
\usepackage{soul}

\usepackage[utf8]{inputenc}
\usepackage[english]{babel}
\usepackage{geometry}
\usepackage{amsmath}
\usepackage{amssymb}
\usepackage{graphicx}

\title{Vaccine allocation to blue-collar workers}
\date{March 2021}

\author[1,2,*]{László Czaller}
\author[1,3]{Gergő Tóth}
\author[1,4]{Balázs Lengyel}
\affil[1]{ELKH Centre for Economic and Regional Studies; Agglomeration and Social Networks Lendület Research Group; Budapest, H-1097, Hungary}
\affil[2]{Eötvös Loránd University; Department for Regional Science; Budapest, H-1117, Hungary}
\affil[3]{University College Dublin; Spatial Dynamics Lab; Belfield, Dublin 4, Ireland}
\affil[4]{Corvinus University of Budapest; Laboratory for Networks, Technology and Innovation; Budapest, H-1093, Hungary}

\affil[*]{Corresponding author: czaller.laszlo@krtk.hu}

\begin{abstract}
Vaccination may be the solution to the pandemic-induced health crisis, but the allocation of vaccines is a complex task in which  economic and social considerations can be important. The central problem is to use the limited number of vaccines in a country to reduce the risk of infection and mitigate economic uncertainty at the same time. In this paper, we propose a simple economic model for vaccine allocation across two types of workers: white-collars can work from home; while blue-collars must work on site. These worker types are complementary to each other, thus a negative shock to the supply of either one decreases the demand for the other that leads to unemployment. Using parameters of blue and white-collar labor supply, their infection risks, productivity losses at home office during lock-down, and available vaccines, we express the optimal share of vaccines allocated to blue-collars. The model points to the dominance of blue-collar vaccination, especially during waves when their relative infection risks increase and when the number of available vaccines is limited. Taking labor supply data from 28 European countries, we quantify blue-collar vaccine allocation that minimizes unemployment across levels of blue- and white-collar infection risks. The model favours blue-collar vaccination identically across European countries in case of vaccine scarcity. As more vaccines become available, economies that host large-shares of employees in home-office shall increasingly immunize them in case blue-collar infection risks can be kept down. Our results highlight that vaccination plans should include workers and rank them by type of occupation. We propose that prioritizing blue-collar workers during infection waves and early vaccination can also favour economy besides helping the most vulnerable who can transmit more infection.

\end{abstract}
\begin{document}

\flushbottom
\maketitle

\thispagestyle{empty}

\section{Introduction}

Humanity has learned in the COVID-19 pandemic that restricting the mobility of individuals and social distancing can slow down virus diffusion \cite{kraemer2020effect, chang2020mobility} but also comes with enormous economic consequences \cite{koren2020business}. Unprecedented fall of demand \cite{guerrieri2020macroeconomic} has led to many job losses \cite{barrero2020covid, kong2020disentangling} and uncertainties in production chains propagate these shocks across countries, sectors, and firms \cite{guan2020global, lenzen2020global}. %,haddad2020structural, bonet2020regional, ivanov2020coronavirus}. 

To balance the mitigation of infection risks with saving economic activity, commuting to work has been a general exception of mobility restrictions even during the most severe quarantines. However, infection risks at work have been found considerable in early outbreak \cite{lan2020work} and remained comparable with other forms of social mixing such as commuting or even nightclubs at later stages \cite{tupper2020event}. Although home-office became the new normal for high-skill and high-income white-collar employees \cite{angelucci2020remote, dingel2020many}, tasks of lower educated workers tend to require physical presence, thus blue-collar workers face higher risks of infection in order to keep their jobs \cite{aum2020should,mongey2020workers}. Higher exposure of certain occupations has been observed through infection tests \cite{de2020occupation}, excess deaths \cite{chen2021excess}, and documented COVID-related deaths \cite{hawkins2021covid}. Partly because income tends to sort white and blue-collar workers to separate neighborhoods, higher exposure is reflected in higher infection rates in densely populated low-income neighborhoods \cite{borjas2020demographic,schmitt2020covid} where the likelihood of within-household infection is also high \cite{malkov2020nature} and where lock-down strategies are less effective \cite{gozzi2020estimating, weill2020social, heroy2021covid}.

Vaccine availability has created a new situation and the question governments must quickly answer is how to locate the limited number of early vaccines to be able to ease restrictions in their country. The consensus is that vaccine allocation must be optimized to save lives and must favor the endangered population \cite{national2020framework, bubar2021model}. However, some argue that certain working groups should be included early in order to speed up vaccination \cite{agarwal2021trade} and immunize those groups that have more contacts and carry more infections \cite{matrajt2020vaccine, forslid2021whom, babus2020optimal, pieroni2021stay}. Despite their obvious importance \cite{manski2021vaccination}, economic rationals are almost completely ignored in this discussion. One exception is Cakmakli et al.\cite{cakmakli2021economic} who illustrate that an ethical distribution of vaccines across countries\cite{emanuel2020ethical} can pay off in functioning global production and supply chains. Yet, the notion that local economies combine white-collar and blue-collar workers differently is still missing from this discourse  and differences across workers are left out from vaccination strategies of most countries.
%grauer: spatiotemporal allocation

In this paper, we build on a fundamental economic argument on complementary tasks of blue- and white-collar workers (also termed routine/non-routine \cite{autor2003skill} and low/high-skill tasks\cite{acemoglu2011skills}) in production on the short-term. This enables us to demonstrate that blue-collar workers should follow the high-risk population in vaccine allocation. Such strategies can favour economies by saving most jobs, besides helping the most vulnerable employees \cite{aum2020should} and mitigating infection transmission \cite{de2020occupation}. 

We propose a short-run economic model in which technology is fixed. % and long-term adjustments to production is limited. 
The model consists of two worker types: white-collars work from home with some productivity losses; while blue-collars work on site. These worker types are used in a fixed combination, thus a negative shock to the supply of either one decreases the demand for the other, leading to unemployment. This model can be used to express the optimal share of vaccines allocated to blue-collars with parameters of labor supply of blue and white-collars, their different infection risks, productivity losses at home office, and the volume of available vaccines. Model results suggest that blue-collars should be prioritized against white-collars during infection waves when the relative infection risks of blue-collars depart from the infection risks of white-collars. This regime of priority is even more pronounced when available vaccines can cover only a small fraction of workers. 

Taking labor supply data from 28 European countries, we quantify blue-collar vaccine allocation that minimizes unemployment across levels of blue- and white-collar infection risks. In case of vaccine availability to only 20\% of all employees, our model minimizes unemployment by allocating 66\% of vaccines to blue-collar workers in 70-80\% of all infection risk scenarios and across all European countries. As more vaccines become available, those European economies that host large-shares of employees in home-office shall increasingly immunize them in case blue-collar infection risks can be kept down. However, economies where blue-collar work dominates, can benefit from continued blue-collar vaccination as more and more vaccines are available, regardless of infection risks.

\section{Model}
\subsection{Basic setup}
Consider an economy, where %representative 
each firm produces a single good by combining two types of tasks: teleworkable tasks ($t$) can be performed from home, and non-teleworkable  tasks ($n$) require the physical presence of workers\cite{dingel2020many, koren2020business}. %respectively. 
Suppose %, to simplify the analysis, 
that workers are able to perform both tasks with unit productivity but once they are trained for one of the tasks, they cannot switch to the other. We assume that the technology combining these tasks is Leontief, that captures short-term production without long-term adjustments to shocks\cite{acemoglu2011skills}, so that the %final good is produced as
aggregate production function is
\begin{equation}\label{1}
    y=\min \left( \alpha_w L_w, \alpha_b L_b  \right)
\end{equation}
where $L_w$ is the amount of white-collar ($w$) labor performing $t$ tasks and and $L_b$ is the amount of blue-collar ($b$) labor performing $n$ tasks, $\alpha_w$ and $\alpha_b$ are unit input requirements and $y$ is aggregate output. Taking labor supply and unit input requirements as given, firms optimize $L_w$ and $L_b$ in order to maximize profit which implies that:  
\begin{equation}\label{2}
    y = \alpha_b L_b = \alpha_w L_w.
\end{equation}

The epidemic starts after firms have optimized labor such that $L_w$ and $L_b$ are fixed. %inputs but before it starts production, . As an element of the usual disease control measures, 
The government obligates firms to send white-collars to home-office. %create the opportunity to work from home in order to reduce the chance of infection among workers. - LB: ne mondjunk ilyet, ha a következő mondatban az szerepel, hogy a blue-collar risk nem változik
As a consequence, the probability of becoming infected will be lower for white-collars %who can perform their jobs remotely 
while the exposure of blue-collar workers performing non-teleworkable tasks will remain unaffected. Formally, let $\beta_i$ be the probability of infection for workers in task $i$ such that $\beta_w \leq \beta_b$. Opportunities for remote work, however, come with a price. Although working from home benefits employees by eliminating their daily commutes it might decrease their productivity by making negotiation, instructing and monitoring more cumbersome, and by increasing reaction time in complex decision situations\cite{koren2020business, bartik2020jobs}. Thus, we assume that the productivity of white-collars decreases to $\gamma \in (0,1)$ as long as they work from home.

Production decreases during the pandemic %the firm produces less 
because effective labor in both tasks deviate negatively from optimal amounts. Without available vaccines the supply of effective labor %in task $t$ and $n$ 
is reduced to
\begin{align*}
   \bar{L}_b = (1 - \beta_b) L_b
\end{align*}
and
\begin{align*}
   \bar{L}_w = (1 - \beta_w) \gamma L_w.
\end{align*}

In the logic of Equation \ref{2}, reduction of effective labor in one task decreases labor demand in the other task and workers become unemployed. Suppose that $\beta_b > 1- (1-\beta_w)\gamma$ holds, so $\beta_b$ reduces the supply of blue-collar labor %in non-teleworkable tasks 
to a greater extent than $\beta_w$ and $\gamma$ decreases white-collar labor. %in teleworkable tasks. 
In this situation, $\left[ (1-\beta_w) \gamma - (1-\beta_b ) \right]L_w$ healthy white-collar workers will be unnecessary for production. If $\beta_b < 1- (1-\beta_w)\gamma$, exactly $\left[(1-\beta_b) - (1-\beta_w)\gamma \right] L_w$ blue-collar workers will become redundant. 

\subsection{Optimal vaccine allocation}
Now suppose that a social planner distributes $V$ amount of vaccines across $L$ workers such that $V < L$. The vaccination has two effects: 1. it immunizes workers and increases effective labor, and 2. it makes social distancing among vaccinated workers unnecessary. After getting the vaccine, white-collar workers performing teleworkable tasks restore their full productivity because they are allowed to go back to the office. Vaccines are scarce and not all workers can be immunized. Therefore, the social planner aims to distribute the vaccines among workers to minimize job losses of healthy workers due to complementary teleworkable and non-teleworkable tasks.%so that as few healthy workers as possible lose their jobs due to the uneven reduction of labor supply in teleworkable and non-teleworkable tasks.
\footnote{Note that the problem of minimizing unemployment among healthy workers and maximizing output provides similar solutions.}

We quantify the share of vaccines that should be allocated to a certain type of worker. Let $v_i$ be the number of immunized workers $i \in \left\{ b,w \right\}$. Assuming that all vaccines are used, $ \sum_{i \in (b,w)} v_i = V $ and every worker accepts the vaccine, labor supplies can be written using $v_b$ only as:
\begin{align}\label{st}
    \bar{L}_b =  (1-\beta_b)L_b + \beta_b v_b,
\end{align}
and
\begin{align}\label{sn}
    \bar{L}_w =  (1-\beta_w)\gamma L_w + (1 - \gamma + \beta_w \gamma)(V-v_b).
\end{align}

The social planner's problem is to minimize the job losses arising from complementary tasks and can be formalized by the objective function
\begin{align*}
    \min_{v_b} \left| \alpha_n \bar{L}_b - \alpha_w \bar{L}_w \right|, \quad 0 \leq v_b \leq V.
\end{align*}
%Clearly, 
The global minimum of the objective function is zero which can be found at 
\begin{align}\label{globmin}
    \underline{v}_b = \frac{\alpha_w \left[ ( 1-\beta_w)\gamma L_w + (1 - \gamma + \beta_w  \gamma )V \right] - (1-\beta_b)\alpha_n L_b}{\alpha_b \beta_b + \alpha_t(1 - \gamma + \beta_w  \gamma )}.
\end{align}
Hence, the solution of the social planner's problem is

\begin{align*}
    v_b^*= \begin{cases}
           0, & \text{if $\underline{v}_b \leq 0 $}\\
           \underline{v}_b, & \text{if $0 < \underline{v}_b < V$ }\\
           V, & \text{if  $\underline{v}_b \geq V.$ } \\
           \end{cases}
\end{align*}

Solutions are depicted in Figure 1. If $\underline{v}_n \leq 0$, we get the corner solution of $v_b^*= 0$, which means that all available vaccines should be given to white-collar workers. %those who perform teleworkable tasks. 
In such cases, the effective white-collar labor determines %the firms' 
output and the total amount of blue-collar work:%labor used for non-teleworkable tasks:
\begin{align*}
    \tilde{L}_b = (1-\beta_w) \gamma L_b + \frac{\alpha_w}{\alpha_b} (1 - \gamma + \beta_w \gamma ) V.
\end{align*}
Substituting $v_b^*=0$ into (Equation \ref{st}), and then subtracting $\tilde{L}_b$ gives the number of unemployed blue-collar workers: %in non-teleworkable tasks:
\begin{align*}
u_b = \left[(1-\beta_b) - (1-\beta_w)\gamma \right] L_w - \frac{\alpha_w}{\alpha_b} (1 - \gamma + \beta_w \gamma ) V.
\end{align*} 
Note that, although this vaccine allocation cannot maintain full employment among healthy blue-collar workers, it still reduces their unemployment due to complementary tasks by $(\alpha_w/\alpha_b) (1 - \gamma + \beta_w \gamma ) V$.

If $\underline{v}_b$ lies within the interval $(0,V)$, the social planner is able to find a vaccination plan that provides the optimal proportion of white-collar and blue-collar workers. %performing teleworkable and non-teleworkable tasks is appropriate.
This implies that nobody drops out of work due to the reduction of work capacity in complementary tasks, so $\tilde{L_i} = \bar{L}_i, \forall i \in (b,w)$.

Finally, if $\underline{v}_b \geq V$, all vaccines should be given to blue-collar workers, $v_b^*= V$, in order to minimize unemployment of white-collar workers. %in teleworkable activities. 
When blue-collar workers are a bottleneck of production, white-collar employment %in non-teleworkable tasks 
becomes
\begin{align*}
    \tilde{L}_w = (1-\beta_w) L_t + \frac{\alpha_b}{\alpha_w} \beta_b V,
\end{align*}
which implies 
\begin{align*}
    u_w = \left[(1-\beta_w) \gamma - (1-\beta_w)\right] L_w - (\alpha_b/\alpha_w) \beta_b V
\end{align*}
unemployed white-collar workers. Compared to the baseline case, this vaccine allocation scheme reduces unemployment in teleworkable tasks by $(\alpha_b/\alpha_w) \beta_b V$.

\begin{figure}[t]
    \centering
    \includegraphics[width = \linewidth]{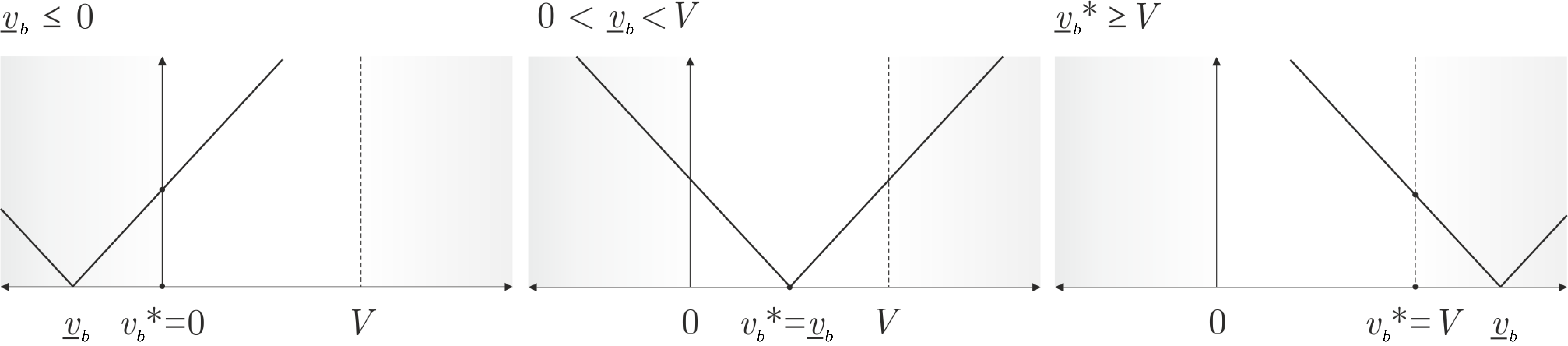}
    \caption{\textbf{Solutions to the social planner's problem.} If $\underline{v}_b \leq 0$ (left), all available vaccines should be given to those who can work from from home. If $0<\underline{v}_b<1$ (middle), $v_b^*=\underline{v}_b$, which means that there is no unemployment among healthy workers. Finally, if $\underline{v}_b \geq V$ all vaccines should be given to blue-collar workers in non-teleworkable tasks.},  
    \label{fig:fig1}
\end{figure} 

The optimal allocation of vaccines depends on blue-collar and white-collar labor supplies ($L_b$ and $L_w$), the parameters describing the structure of the economy ($\alpha_b$, $\alpha_w$, and $\gamma$), the number of vaccines available ($V$) and the task-specific probabilities of infection ($\beta_b$ and $\beta_w$). It follows that there is no uniform recipe for the distribution of vaccines which derives solely from the characteristics of the economy. By differentiating equation (\ref{globmin}) with respect to $\beta_b$ we obtain
\begin{align*}
    \frac{\partial \underline{v}_b}{\partial \beta_b} = \frac{\alpha_b}{\alpha_b \beta_b + \alpha_w(1 - \gamma + \beta_w  \gamma )} \left( L_b - \underline{v}_b \right),
\end{align*}
hence if the probability of infection for blue-collar workers performing non-teleworkable tasks increases (e.g. safety regulations are not followed by employees at the workplace), more vaccines should be given to them not just to avoid workplace infection but also to minimize redundancy of white-collar labor. Clearly, if all blue-collar workers are vaccinated, $\underline{v}_b = L_b$, $\beta_b$ has no further effects on $\underline{v}_b$. 

In contrast, by differentiating $\underline{v}_b$ with respect to $\beta_w$ we find that an infinitesimal change in the infection risk of white-collar workers decreases the optimal number of vaccinated blue-collar workers ($\underline{v}_b$),
\begin{align*}
    \frac{\partial \underline{v}_b}{\partial \beta_w} = \frac{\alpha_w \gamma}{\alpha_b \beta_b + \alpha_w(1 - \gamma + \beta_w  \gamma )} \left( V - L_w - \underline{v}_b \right),
\end{align*}
as long as $\underline{v}_w < L_w$. 
\section{Results}
We illustrate the results of our model with numerical simulations. We relate our theoretical results to pre-Covid-19 observational data to examine the extent to which $v_b^*$ varies across economies of different structures and various epidemiological scenarios represented by different combinations of theoretical infection risks $\beta_b$ and $\beta_w$ that range from 0 (no risk of infection) to 1 (infection is certain). We collect employment data from the EU Labor Force Survey\footnote{Source: https://ec.europa.eu/eurostat/web/lfs/data/database} for countries of the European Economic Area (EEA). To calculate $L_w$ and $L_b$ for each country, we use the country-level estimates of Dingel and Neiman on the share of jobs that can be done at home.\cite{dingel2020many} We calibrate unit input requirements to match model equations to employment data. First, $\alpha_w$ is normalized to unity and $\alpha_b$ is estimated by substituting $L_w$ and $L_b$ into Equation \ref{2}. Finally, remote work productivity $\gamma$ is set to $0.8$ following the results of a recent survey.\cite{bartik2020jobs} 

\begin{figure}[h]
    \centering
    \includegraphics[width = 0.75\linewidth]{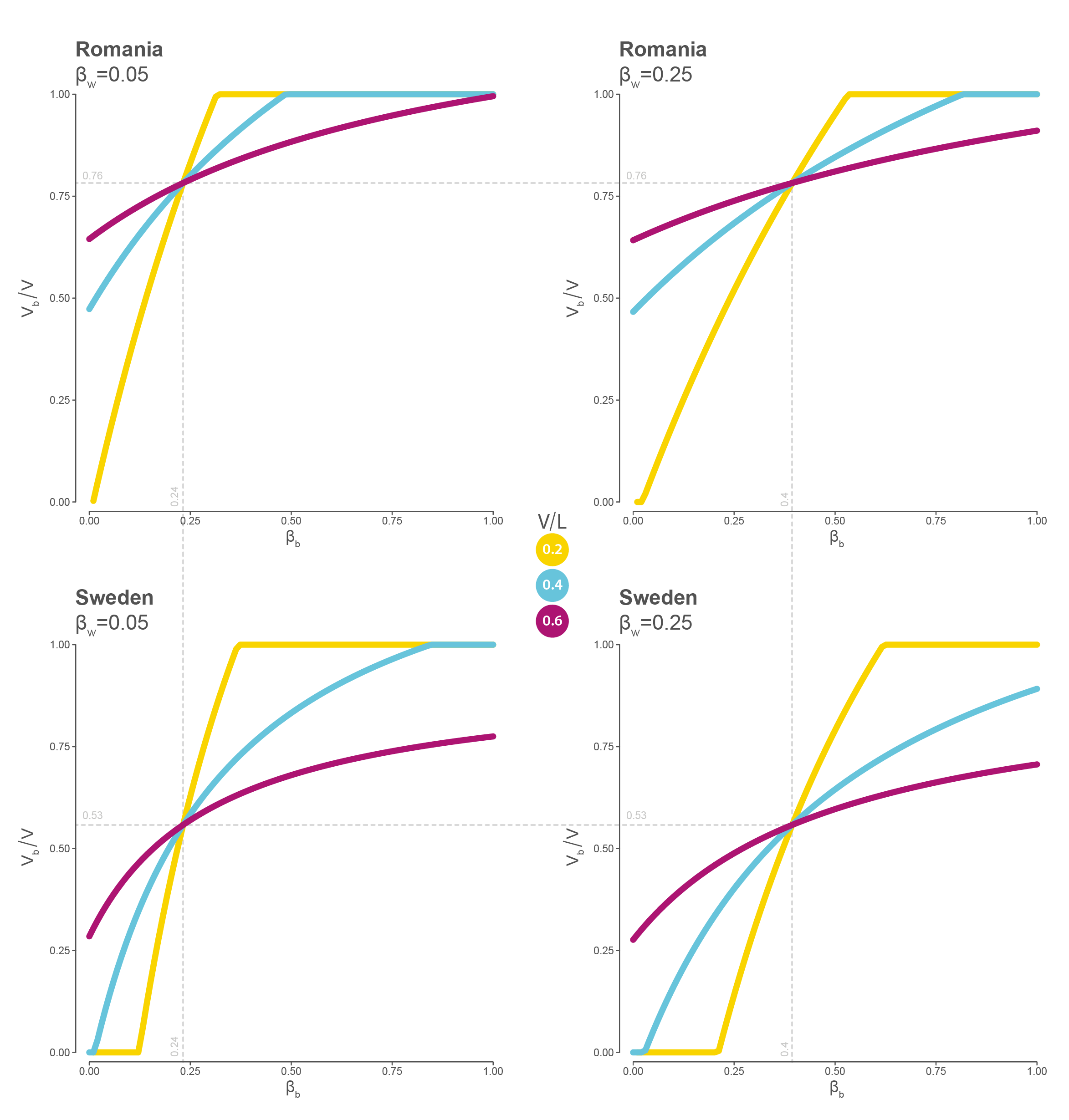}
     \caption{\textbf{Optimal blue-collar vaccination as a function of blue-collar infection risk and available vaccines in two example countries.} We fix white-collar infection risks at $\beta_w=0.05$ (left column) and $\beta_w=0.25$ (right column). Line colors refer to vaccine availability $V/L=0.2$ yellow, $V/L=0.4$ blue, $V/L=0.6$ purple.} 
     \label{fig:fig2}
\end{figure} 

Figure \ref{fig:fig2} depicts the share of vaccines given to blue-collar workers ($v_b/V$) as a function of $\beta_b$ for two countries with different shares of blue-collar jobs: Sweden (low $L_b/L$), %\hl{Germany (average $L_w/L$)} % if it was yellow to be checked whether it's indeed the average, then it is
and Romania (high $L_b/L$). Each subplot considers three scenarios that represent different levels of vaccine scarcity. For example, the first scenario described by $V/L = 0.2$ (yellow) means that only 20\% of employees receive a vaccine while the scenario of $V/L = 0.6$ (purple) describes the case when there is enough vaccines to immunize 60\% of the workers. 

\begin{figure}[!b]
    \centering
    \includegraphics[width = 0.65\linewidth]{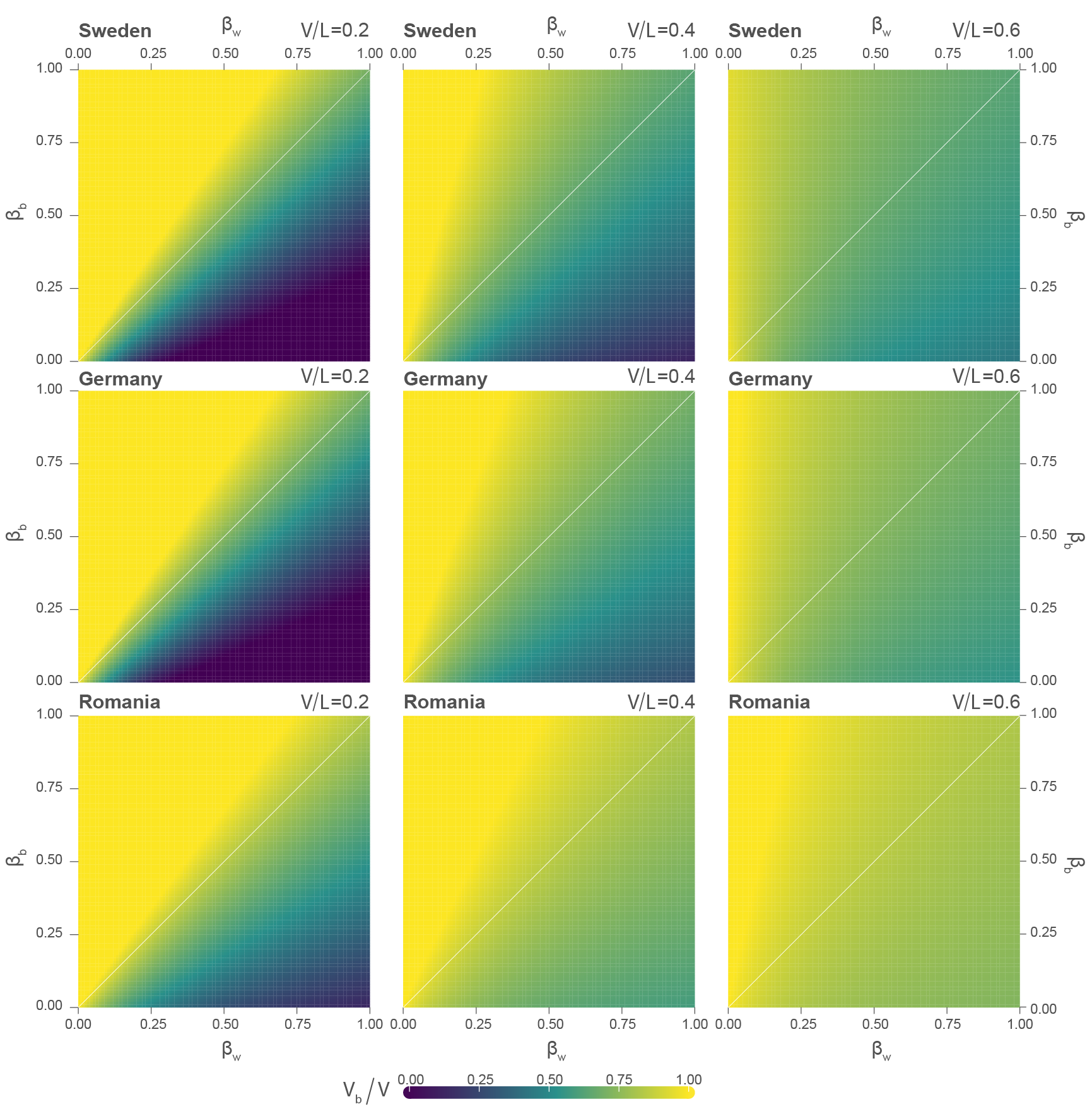}
     \caption{\textbf{Vaccine allocation by occupation-specific infection risks and available vaccines.} Colors represent values of blue-collar fraction in vaccines $v_b$, $\beta_w$ is theoretical infection risk of white-collars, $\beta_b$ is theoretical infection risk of blue-collars. $V/L$ stand for the share of worker population that can be vaccinated with available vaccines.}
     \label{fig:fig3}
\end{figure} 

As expected, $v_b/V$ increases in $\beta_b$ irrespective of the choice of $\beta_w$, however, the curves characterizing the relationship between the two variables vary considerably across scenarios. In the case of extreme vaccine scarcity ($V/L = 0.2$), $v_b/V$ is low when $\beta_b$ is similar to $\beta_w$, $v_b/V$ increases sharply with $\beta_b$ and peaks quickly in each country demonstrating the urgent need of immunizing blue-collar workers to avoid production losses caused by relatively high infection risks of working on site. 
In case $V/L$ is higher, the curves start at higher $v_b/V$ values and have smoother slopes demonstrating %that the intensive vaccination of blue-collar workers should be accompanied by the immunization of those working from home, especially when $\beta_w$ is higher, as vaccine availability improves. - LB: disagree. the slope refers to beta_b elasticity, which is infection wave 
that higher fraction of vaccines shall be allocated to blue-collar workers even at small infection rates but improvements of vaccine availability decreases the influence of infection waves when $\beta_b$ exceeds $\beta_w$. Countries differ in terms of their $L_b/L$ ratio; thus, require different $v_b/V$. % In case of high $V$, blue-collar vaccination must start from higher $v_b/V$ in Romania than in Sweden. 

An interesting feature of the model is the intersection of curves at $\beta_b = 1 - (1-\beta_w)\gamma$. In this point, redundancies of blue-collar and white-collar labor are identical, regardless of $V$. %This condition describes the situation that the amount of labor lost from type-$b$ tasks is equal to the amount of labor dropping out from tasks that can be done at home. 
These intersections capture the level of optimal blue-collar vaccination that the country converges to as more and more vaccines are available. If $\beta_b > 1 - (1-\beta_w)\gamma$, infection of blue-collar workers can generate redundancy of white-collar labor and thus blue-collar vaccination shall be prioritized when $V/L$ is low. 
%more labor is missed from non-teleworkable tasks due the infection therefore greater emphasis should be placed on the immunization of blue-collar workers when $V/L$ is low in order to avoid unemployment among healthy workers in type-$w$ tasks. 
In contrast, $\beta_b < 1 - (1-\beta_w)\gamma$ %, the amount of labor missing from teleworkable tasks is higher which 
implies low $v_b/V$ values in scenarios where $V/L$ is low because white-collar labor losses cause redundancies in blue-collar jobs. Curves converge to the $v_b/V$ value at their intersection such that blue-collar vaccination increases with $V$ in the $\beta_b < 1 - (1-\beta_w)\gamma$ regime but increasing $V$ drops $v_b/V$ in the $\beta_b > 1 - (1-\beta_w)\gamma$ regime.
%Since countries differ in terms of $L_w/L$ ratios, If $L_b/L$ is relatively low, the optimal vaccination plan should imply low $v_b/V$ ratios at any given $\beta_b$ infection risk. For example, in Sweden, where the share of teleworkable jobs is higher than in Germany or Romania, more vaccines should be given to those who can work from home. Otherwise, the over-vaccination of blue-collar workers would result in redundant labor in \hl{type-$w$} % sure not type b?
%tasks.
% I would frame it from the the other way around. It doesn't look sexy to me for arguing vaccination for white-collar workers. I'd go like: - LB: I vote for Gergő's version but reframe it to vaccine availability
The convergent $v_b/V$ level is country specific and depends on the country's $L_b/L$ ratio. If $L_b/L$ is relatively high, improving vaccine availability designates high $v_b/V$ ratios at any given $\beta_n$ infection risk. For example, in Romania, where the share of non-teleworkable jobs is higher than in %Germany or 
Sweden, more vaccines should be given to blue-collar workers who can not work from home. Otherwise, the over-vaccination of white-collar workers would result in redundant blue-collar labor.

So far, we have analysed the relationship between $v_b/V$ and $\beta_b$ at given $\beta_w$ infection probabilities. Now, we proceed by calculating the share of vaccines given to blue-collar workers for all possible combinations of $\beta_w$ and $\beta_b$. Figure \ref{fig:fig3} organizes $v_b/V$ ratios into matrices for Sweden (low $L_b/L$) , Germany (medium $L_b/L$), and Romania (high $L_b/L$) at different levels of vaccine scarcity. Perhaps the most conspicuous pattern of this figure is that in countries where the rate of remote work is high (e.g. Sweden and Germany), more epidemic scenarios can be found where optimal vaccine allocations favor individuals performing teleworkable tasks. However, in countries where remote work is less common (see e.g. Romania) high $v_b/V$ ratios can be observed even in cases where $\beta_w$ exceeds $\beta_b$. If we consider only those parameter combinations where $\beta_b > \beta_w$ (above the matrix diagonal), differences between countries are blurred, especially when vaccines are widely available. Taken together, these results suggest that blue-collar vaccination should exceed white-collar vaccination regardless of infection risks and vaccine availability. %On this basis, a higher proportion of available vaccines should be given to individuals who, due to the nature of their jobs, cannot work from home. 

Since task-specific infection probabilities, as they appear in the model, are extremely difficult to relate to observational data, exact values for $v_b/V$ cannot be given. This would not make much sense anyway, because $\beta_w$ and $\beta_b$ may vary depending on the phases of the epidemic and the protective measures imposed. %\hl[For example, the absence of protective measures and restrictions result in higher infection risks in general.] 
However, matrices in Figure \ref{fig:fig3} can be used to provide a crude measure that informs us about the need to vaccinate blue-collar workers in various epidemiological scenarios. For this, we calculate the percentage share of $\beta_b > \beta_w$ infection risk combinations where $v_b/V$ exceeds a $v_b^{th}$ threshold. %to the total number of combinations where $\beta_b > \beta_w$. 
For example, by setting $v_b^{th}$ to 0.66, we calculate the proportion of combinations where at least two-thirds of the vaccines are given to blue-collar workers. The higher the ratio obtained, the more important it is to vaccinate blue-collar workers. 

\begin{figure}[!b]
    \centering
    \includegraphics[width = 0.8\linewidth]{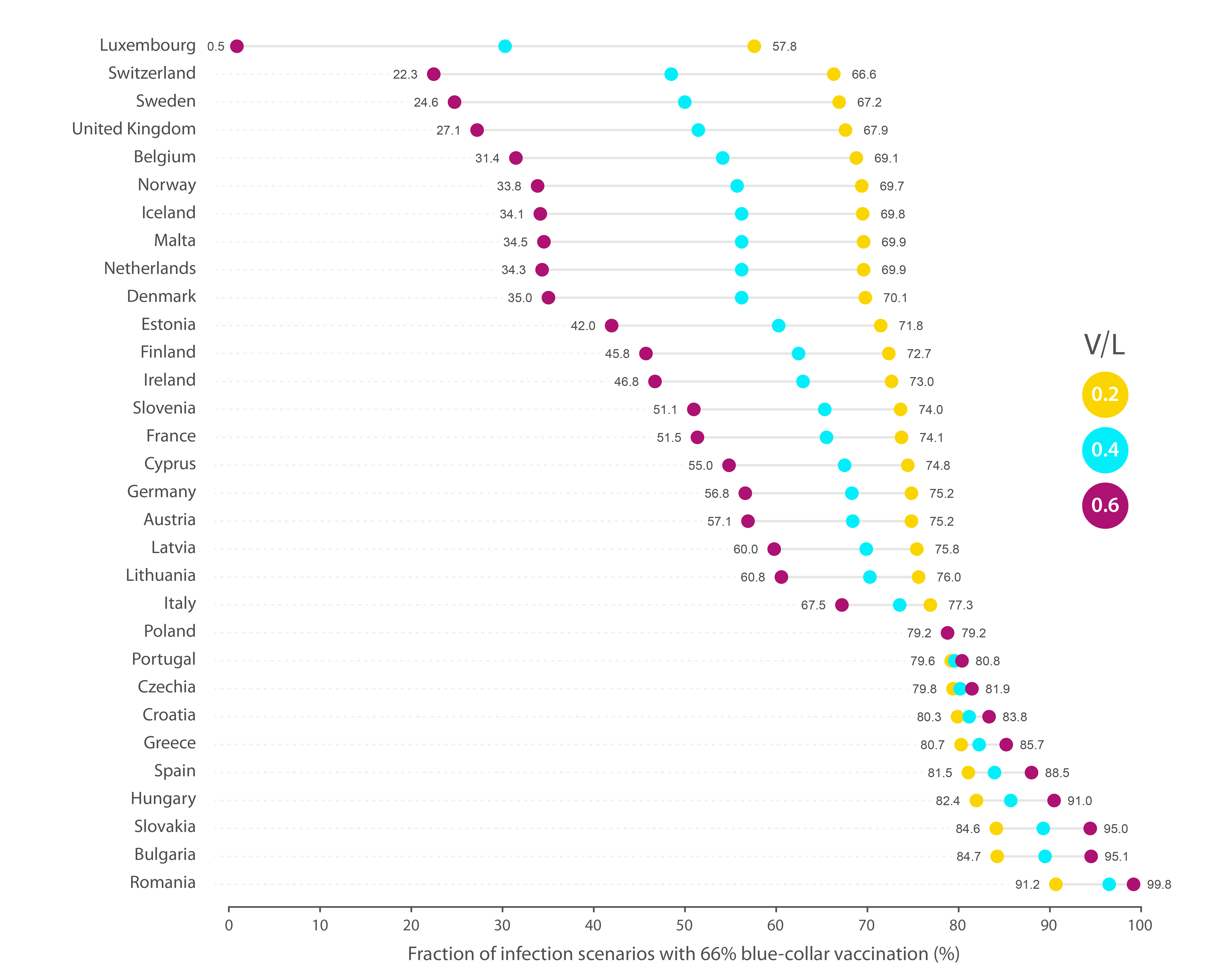}
     \caption{\textbf{Blue-collar vaccination saves jobs in European countries across infections risks and vaccine availability.} $V/L$ denotes vaccine availability for 20-40-60\% of all employees. Percentage values represent the share of infection scenarios when blue-collar workers shall receive more than 66\% of available vaccines to save the most jobs.}
     \label{fig:fig4}
\end{figure} 

Figure \ref{fig:fig4} shows the ratio of those $\beta_b > \beta_w$ combinations where $v_b/V > 0.66$. In case of serious vaccine shortage ($V/L=0.2$), unemployment can be minimized by allocating at least 66\% of vaccines to blue-collar workers in 70-80\% of the considered epidemic scenarios in most European countries. As more vaccines become available, countries where remote work is more prevalent (e.g. Luxembourg, Switzerland and Sweden) shall start to vaccinate teleworkers in larger proportions to ensure full employment among healthy workers. These countries should aim for a more balanced vaccine allocation unless $\beta_b$ is too high compared to $\beta_w$ (see the Swedish case in Figure \ref{fig:fig3}.) However, in several countries where the share of teleworkable jobs is low (e.g. Romania, Bulgaria, Slovakia), mass vaccination of blue-collars remains necessary even if vaccines become widely available. 

Clearly, the lower $v_b^{th}$ the less variation can be observed across countries. This is due to the general notion illustrated in Figure \ref{fig:fig3} that when $\beta_b > \beta_w$ blue-collar workers should almost always get more vaccines. For example, if $v_b^{th}$ is set to $0.5$ 80-90\% of the relevant risk combinations satisfy the condition $v_b/V > v_b^{th}$ in all countries except Luxembourg (illustrated in Supporting Figure \ref{fig:SI_fig1}). However, the convergent $v_b/V$ nature of increasing vaccine availability illustrated in Figure \ref{fig:fig2} implies that countries of low $L_b/L$ (eg. Sweden) can allocate vaccines to white-collars earlier than countries of high $L_b/L$ (eg. Romania).

\section{Discussion}

Vaccinating those who must go out to work even during the most sever phases of the pandemic can save lives. Only very few European countries, like Ireland and to some extent Spain\cite{pieroni2021stay}, differentiate occupations in their vaccination plans, but this practice remains exception and it is certainly not the rule yet. Our paper shows that immunization of blue-collar workers who can't work at home can save jobs as well. This is done in an abstract model by assuming that teleworkable and non-teleworkable tasks are complements in the production on the short run. Thus, prioritizing blue-collars in the vaccination strategies can be beneficial for white-collar jobs as well.

We quantify optimal shares of vaccines allocated to blue-collars by minimizing the unemployment arising from infection risks and productivity losses at home office.
Our model suggests that European countries should allocate majority of the early vaccines for workers to blue-collars. 
%In those more developed European economies that host large shares of home-office jobs, white-collars can be increasingly vaccinated after 50\% of blue-collars have been immunized, if blue-collar infection risks are high. However, in less developed countries, where blue-collar work is more dominant, white-collar vaccination should come later, at 70\% of blue-collar immunization.
As more vaccines become available, more advanced should increasingly immunize white-collars working in home-office in case blue-collar infection risks can be kept down. However, in less developed economies where blue-collar work dominates, should focus more on blue-collar vaccination.

Further work is needed to overcome the limitations of the model presented here. We are assuming that economic agents are not adjusting themselves to the dynamically changing circumstances over the pandemic job losses. Home office productivity losses are assumed to be identical across countries, which is probably not true given the technological and cultural differences that might influence the efficiency of working at home. Epidemic scenarios are completely ignored in this framework and infection risks are only considered in a very abstract manner. Finally, production only considers labor while other inputs and demand are ignored.

%Total stock of labor in teleworkable and non-teleworkable jobs for all EU member states we use aggregate employment data from the European Union's Labor Force Survey (EU LFS) and country-level estimates on the share of teleworkable jobs. These estimates come form the seminal work of Dingel and Neiman who classified occupations according to their feasibility of remote work and then by applying this classification, provided estimates on the share of teleworkable jobs for 85 countries \hl{[CITE Dingel, Neiman]}.  input requirements a calibrated 

\typeout{}
\bibliography{references}

\begin{thebibliography}{10}
\urlstyle{rm}
\expandafter\ifx\csname url\endcsname\relax
  \def\url#1{\texttt{#1}}\fi
\expandafter\ifx\csname urlprefix\endcsname\relax\def\urlprefix{URL }\fi
\expandafter\ifx\csname doiprefix\endcsname\relax\def\doiprefix{DOI: }\fi
\providecommand{\bibinfo}[2]{#2}
\providecommand{\eprint}[2][]{\url{#2}}

\bibitem{kraemer2020effect}
\bibinfo{author}{Kraemer, M.~U.} \emph{et~al.}
\newblock \bibinfo{journal}{\bibinfo{title}{The effect of human mobility and
  control measures on the covid-19 epidemic in china}}.
\newblock {\emph{\JournalTitle{Science}}} \textbf{\bibinfo{volume}{368}},
  \bibinfo{pages}{493--497} (\bibinfo{year}{2020}).

\bibitem{chang2020mobility}
\bibinfo{author}{Chang, S.} \emph{et~al.}
\newblock \bibinfo{journal}{\bibinfo{title}{Mobility network models of covid-19
  explain inequities and inform reopening}}.
\newblock {\emph{\JournalTitle{Nature}}} \bibinfo{pages}{1--6}
  (\bibinfo{year}{2020}).

\bibitem{koren2020business}
\bibinfo{author}{Koren, M.} \& \bibinfo{author}{Pet{\H{o}}, R.}
\newblock \bibinfo{journal}{\bibinfo{title}{Business disruptions from social
  distancing}}.
\newblock {\emph{\JournalTitle{Plos one}}} \textbf{\bibinfo{volume}{15}},
  \bibinfo{pages}{e0239113} (\bibinfo{year}{2020}).

\bibitem{guerrieri2020macroeconomic}
\bibinfo{author}{Guerrieri, V.}, \bibinfo{author}{Lorenzoni, G.},
  \bibinfo{author}{Straub, L.} \& \bibinfo{author}{Werning, I.}
\newblock \bibinfo{title}{Macroeconomic implications of covid-19: Can negative
  supply shocks cause demand shortages?}
\newblock \bibinfo{type}{Tech. Rep.}, \bibinfo{institution}{National Bureau of
  Economic Research} (\bibinfo{year}{2020}).

\bibitem{barrero2020covid}
\bibinfo{author}{Barrero, J.~M.}, \bibinfo{author}{Bloom, N.} \&
  \bibinfo{author}{Davis, S.~J.}
\newblock \bibinfo{title}{Covid-19 is also a reallocation shock}.
\newblock \bibinfo{type}{Tech. Rep.}, \bibinfo{institution}{National Bureau of
  Economic Research} (\bibinfo{year}{2020}).

\bibitem{kong2020disentangling}
\bibinfo{author}{Kong, E.} \& \bibinfo{author}{Prinz, D.}
\newblock \bibinfo{journal}{\bibinfo{title}{Disentangling policy effects using
  proxy data: Which shutdown policies affected unemployment during the covid-19
  pandemic?}}
\newblock {\emph{\JournalTitle{Journal of Public Economics}}}
  \textbf{\bibinfo{volume}{189}}, \bibinfo{pages}{104257}
  (\bibinfo{year}{2020}).

\bibitem{guan2020global}
\bibinfo{author}{Guan, D.} \emph{et~al.}
\newblock \bibinfo{journal}{\bibinfo{title}{Global supply-chain effects of
  covid-19 control measures}}.
\newblock {\emph{\JournalTitle{Nature Human Behaviour}}} \bibinfo{pages}{1--11}
  (\bibinfo{year}{2020}).

\bibitem{lenzen2020global}
\bibinfo{author}{Lenzen, M.} \emph{et~al.}
\newblock \bibinfo{journal}{\bibinfo{title}{Global socio-economic losses and
  environmental gains from the coronavirus pandemic}}.
\newblock {\emph{\JournalTitle{PLoS One}}} \textbf{\bibinfo{volume}{15}},
  \bibinfo{pages}{e0235654} (\bibinfo{year}{2020}).

\bibitem{lan2020work}
\bibinfo{author}{Lan, F.-Y.}, \bibinfo{author}{Wei, C.-F.},
  \bibinfo{author}{Hsu, Y.-T.}, \bibinfo{author}{Christiani, D.~C.} \&
  \bibinfo{author}{Kales, S.~N.}
\newblock \bibinfo{journal}{\bibinfo{title}{Work-related covid-19
  transmission}}.
\newblock {\emph{\JournalTitle{medRxiv}}}  (\bibinfo{year}{2020}).

\bibitem{tupper2020event}
\bibinfo{author}{Tupper, P.}, \bibinfo{author}{Boury, H.},
  \bibinfo{author}{Yerlanov, M.} \& \bibinfo{author}{Colijn, C.}
\newblock \bibinfo{journal}{\bibinfo{title}{Event-specific interventions to
  minimize covid-19 transmission}}.
\newblock {\emph{\JournalTitle{Proceedings of the National Academy of
  Sciences}}} \textbf{\bibinfo{volume}{117}}, \bibinfo{pages}{32038--32045}
  (\bibinfo{year}{2020}).

\bibitem{angelucci2020remote}
\bibinfo{author}{Angelucci, M.}, \bibinfo{author}{Angrisani, M.},
  \bibinfo{author}{Bennett, D.~M.}, \bibinfo{author}{Kapteyn, A.} \&
  \bibinfo{author}{Schaner, S.~G.}
\newblock \bibinfo{title}{Remote work and the heterogeneous impact of covid-19
  on employment and health}.
\newblock \bibinfo{type}{Tech. Rep.}, \bibinfo{institution}{National Bureau of
  Economic Research} (\bibinfo{year}{2020}).

\bibitem{dingel2020many}
\bibinfo{author}{Dingel, J.~I.} \& \bibinfo{author}{Neiman, B.}
\newblock \bibinfo{title}{How many jobs can be done at home?}
\newblock \bibinfo{type}{Tech. Rep.}, \bibinfo{institution}{National Bureau of
  Economic Research} (\bibinfo{year}{2020}).

\bibitem{aum2020should}
\bibinfo{author}{Aum, S.}, \bibinfo{author}{Lee, S. Y.~T.} \&
  \bibinfo{author}{Shin, Y.}
\newblock \bibinfo{title}{Who should work from home during a pandemic? the
  wage-infection trade-off}.
\newblock \bibinfo{type}{Tech. Rep.}, \bibinfo{institution}{National Bureau of
  Economic Research} (\bibinfo{year}{2020}).

\bibitem{mongey2020workers}
\bibinfo{author}{Mongey, S.}, \bibinfo{author}{Pilossoph, L.} \&
  \bibinfo{author}{Weinberg, A.}
\newblock \bibinfo{title}{Which workers bear the burden of social distancing
  policies? nber working paper 27085, may}.
\newblock \bibinfo{type}{Tech. Rep.} (\bibinfo{year}{2020}).

\bibitem{de2020occupation}
\bibinfo{author}{de~Gier, B.} \emph{et~al.}
\newblock \bibinfo{journal}{\bibinfo{title}{Occupation-and age-associated risk
  of sars-cov-2 test positivity, the netherlands, june to october 2020}}.
\newblock {\emph{\JournalTitle{Eurosurveillance}}}
  \textbf{\bibinfo{volume}{25}}, \bibinfo{pages}{2001884}
  (\bibinfo{year}{2020}).

\bibitem{chen2021excess}
\bibinfo{author}{Chen, Y.-H.} \emph{et~al.}
\newblock \bibinfo{journal}{\bibinfo{title}{Excess mortality associated with
  the covid-19 pandemic among californians 18-65 years of age, by occupational
  sector and occupation: March through october 2020}}.
\newblock {\emph{\JournalTitle{medRxiv}}}  (\bibinfo{year}{2021}).

\bibitem{hawkins2021covid}
\bibinfo{author}{Hawkins, D.}, \bibinfo{author}{Davis, L.} \&
  \bibinfo{author}{Kriebel, D.}
\newblock \bibinfo{journal}{\bibinfo{title}{Covid-19 deaths by occupation,
  massachusetts, march 1--july 31, 2020}}.
\newblock {\emph{\JournalTitle{American journal of industrial medicine}}}
  (\bibinfo{year}{2021}).

\bibitem{borjas2020demographic}
\bibinfo{author}{Borjas, G.~J.}
\newblock \bibinfo{title}{Demographic determinants of testing incidence and
  covid-19 infections in new york city neighborhoods}.
\newblock \bibinfo{type}{Tech. Rep.}, \bibinfo{institution}{National Bureau of
  Economic Research} (\bibinfo{year}{2020}).

\bibitem{schmitt2020covid}
\bibinfo{author}{Schmitt-Groh{\'e}, S.}, \bibinfo{author}{Teoh, K.} \&
  \bibinfo{author}{Uribe, M.}
\newblock \bibinfo{title}{Covid-19: Testing inequality in new york city}.
\newblock \bibinfo{type}{Tech. Rep.}, \bibinfo{institution}{National Bureau of
  Economic Research} (\bibinfo{year}{2020}).

\bibitem{malkov2020nature}
\bibinfo{author}{Malkov, E.}
\newblock \bibinfo{journal}{\bibinfo{title}{Nature of work and distribution of
  risk: Evidence from occupational sorting, skills, and tasks}}.
\newblock {\emph{\JournalTitle{CEPR Covid Economics: Vetted and Real Time
  Papers}}} \textbf{\bibinfo{volume}{34}}, \bibinfo{pages}{15--49}
  (\bibinfo{year}{2020}).

\bibitem{gozzi2020estimating}
\bibinfo{author}{Gozzi, N.} \emph{et~al.}
\newblock \bibinfo{journal}{\bibinfo{title}{Estimating the effect of social
  inequalities in the mitigation of covid-19 across communities in santiago de
  chile}}.
\newblock {\emph{\JournalTitle{medRxiv}}}  (\bibinfo{year}{2020}).

\bibitem{weill2020social}
\bibinfo{author}{Weill, J.~A.}, \bibinfo{author}{Stigler, M.},
  \bibinfo{author}{Deschenes, O.} \& \bibinfo{author}{Springborn, M.~R.}
\newblock \bibinfo{journal}{\bibinfo{title}{Social distancing responses to
  covid-19 emergency declarations strongly differentiated by income}}.
\newblock {\emph{\JournalTitle{Proceedings of the National Academy of
  Sciences}}} \textbf{\bibinfo{volume}{117}}, \bibinfo{pages}{19658--19660}
  (\bibinfo{year}{2020}).

\bibitem{heroy2021covid}
\bibinfo{author}{Heroy, S.}, \bibinfo{author}{Loaiza, I.},
  \bibinfo{author}{Pentland, A.} \& \bibinfo{author}{O’Clery, N.}
\newblock \bibinfo{journal}{\bibinfo{title}{Covid-19 policy analysis: labour
  structure dictates lockdown mobility behaviour}}.
\newblock {\emph{\JournalTitle{Journal of the Royal Society Interface}}}
  \textbf{\bibinfo{volume}{18}}, \bibinfo{pages}{20201035}
  (\bibinfo{year}{2021}).

\bibitem{national2020framework}
\bibinfo{author}{National Academies~of Sciences, E.},
  \bibinfo{author}{Medicine} \emph{et~al.}
\newblock \emph{\bibinfo{title}{Framework for equitable allocation of COVID-19
  vaccine}} (\bibinfo{publisher}{National Academies Press},
  \bibinfo{year}{2020}).

\bibitem{bubar2021model}
\bibinfo{author}{Bubar, K.~M.} \emph{et~al.}
\newblock \bibinfo{journal}{\bibinfo{title}{Model-informed covid-19 vaccine
  prioritization strategies by age and serostatus}}.
\newblock {\emph{\JournalTitle{Science}}} \textbf{\bibinfo{volume}{371}},
  \bibinfo{pages}{916--921} (\bibinfo{year}{2021}).

\bibitem{agarwal2021trade}
\bibinfo{author}{Agarwal, N.}, \bibinfo{author}{Komo, A.},
  \bibinfo{author}{Patel, C.~A.}, \bibinfo{author}{Pathak, P.~A.} \&
  \bibinfo{author}{{\"U}nver, M.~U.}
\newblock \bibinfo{title}{The trade-off between prioritization and vaccination
  speed depends on mitigation measures}.
\newblock \bibinfo{type}{Tech. Rep.}, \bibinfo{institution}{National Bureau of
  Economic Research} (\bibinfo{year}{2021}).

\bibitem{matrajt2020vaccine}
\bibinfo{author}{Matrajt, L.}, \bibinfo{author}{Eaton, J.},
  \bibinfo{author}{Leung, T.} \& \bibinfo{author}{Brown, E.~R.}
\newblock \bibinfo{journal}{\bibinfo{title}{Vaccine optimization for covid-19:
  who to vaccinate first?}}
\newblock {\emph{\JournalTitle{Science Advances}}}  (\bibinfo{year}{2021}).

\bibitem{forslid2021whom}
\bibinfo{author}{Forslid, R.} \& \bibinfo{author}{Herzing, M.}
\newblock \bibinfo{title}{Whom to vaccinate first-some important trade-offs}.
\newblock \bibinfo{type}{Tech. Rep.} (\bibinfo{year}{2021}).

\bibitem{babus2020optimal}
\bibinfo{author}{Babus, A.}, \bibinfo{author}{Das, S.} \& \bibinfo{author}{Lee,
  S.}
\newblock \bibinfo{journal}{\bibinfo{title}{The optimal allocation of covid-19
  vaccines}}.
\newblock {\emph{\JournalTitle{medRxiv}}}  (\bibinfo{year}{2020}).

\bibitem{pieroni2021stay}
\bibinfo{author}{Pieroni, V.}, \bibinfo{author}{Facchini, A.} \&
  \bibinfo{author}{Riccaboni, M.}
\newblock \bibinfo{journal}{\bibinfo{title}{From stay-at-home to return-to-work
  policies: Covid-19 mortality, mobility and furlough schemes in italy}}.
\newblock {\emph{\JournalTitle{arXiv preprint arXiv:2102.03619}}}
  (\bibinfo{year}{2021}).

\bibitem{manski2021vaccination}
\bibinfo{author}{Manski, C.~F.}
\newblock \bibinfo{title}{Vaccination planning under uncertainty, with
  application to covid-19}.
\newblock \bibinfo{type}{Tech. Rep.}, \bibinfo{institution}{National Bureau of
  Economic Research} (\bibinfo{year}{2021}).

\bibitem{cakmakli2021economic}
\bibinfo{author}{Cakmakli, C.}, \bibinfo{author}{Demiralp, S.},
  \bibinfo{author}{Kalemli-{\"O}zcan, S.}, \bibinfo{author}{Yildirim, M.~A.}
  \emph{et~al.}
\newblock \bibinfo{title}{The economic case for global vaccinations: an
  epidemiological model with international production networks}.
\newblock \bibinfo{type}{Tech. Rep.} (\bibinfo{year}{2021}).

\bibitem{emanuel2020ethical}
\bibinfo{author}{Emanuel, E.~J.} \emph{et~al.}
\newblock \bibinfo{journal}{\bibinfo{title}{An ethical framework for global
  vaccine allocation}}.
\newblock {\emph{\JournalTitle{Science}}} \textbf{\bibinfo{volume}{369}},
  \bibinfo{pages}{1309--1312} (\bibinfo{year}{2020}).

\bibitem{autor2003skill}
\bibinfo{author}{Autor, D.~H.}, \bibinfo{author}{Levy, F.} \&
  \bibinfo{author}{Murnane, R.~J.}
\newblock \bibinfo{journal}{\bibinfo{title}{The skill content of recent
  technological change: An empirical exploration}}.
\newblock {\emph{\JournalTitle{The Quarterly journal of economics}}}
  \textbf{\bibinfo{volume}{118}}, \bibinfo{pages}{1279--1333}
  (\bibinfo{year}{2003}).

\bibitem{acemoglu2011skills}
\bibinfo{author}{Acemoglu, D.} \& \bibinfo{author}{Autor, D.}
\newblock \bibinfo{title}{Skills, tasks and technologies: Implications for
  employment and earnings}.
\newblock In \emph{\bibinfo{booktitle}{Handbook of labor economics}},
  vol.~\bibinfo{volume}{4}, \bibinfo{pages}{1043--1171}
  (\bibinfo{publisher}{Elsevier}, \bibinfo{year}{2011}).

\bibitem{bartik2020jobs}
\bibinfo{author}{Bartik, A.~W.}, \bibinfo{author}{Cullen, Z.~B.},
  \bibinfo{author}{Glaeser, E.~L.}, \bibinfo{author}{Luca, M.} \&
  \bibinfo{author}{Stanton, C.~T.}
\newblock \bibinfo{title}{What jobs are being done at home during the covid-19
  crisis? evidence from firm-level surveys}.
\newblock \bibinfo{type}{Tech. Rep.}, \bibinfo{institution}{National Bureau of
  Economic Research} (\bibinfo{year}{2020}).

\end{thebibliography}

\renewcommand{\figurename}{Supplementary Figure}
\renewcommand{\tablename}{Supplementary Table}
\renewcommand{\theequation}{S\arabic{equation}} 
\setcounter{figure}{0}
\setcounter{table}{0}
\setcounter{equation}{0}

\clearpage

\section*{Appendix}
\begin{figure}[h]
    \centering
    \includegraphics[width = 0.9\linewidth]{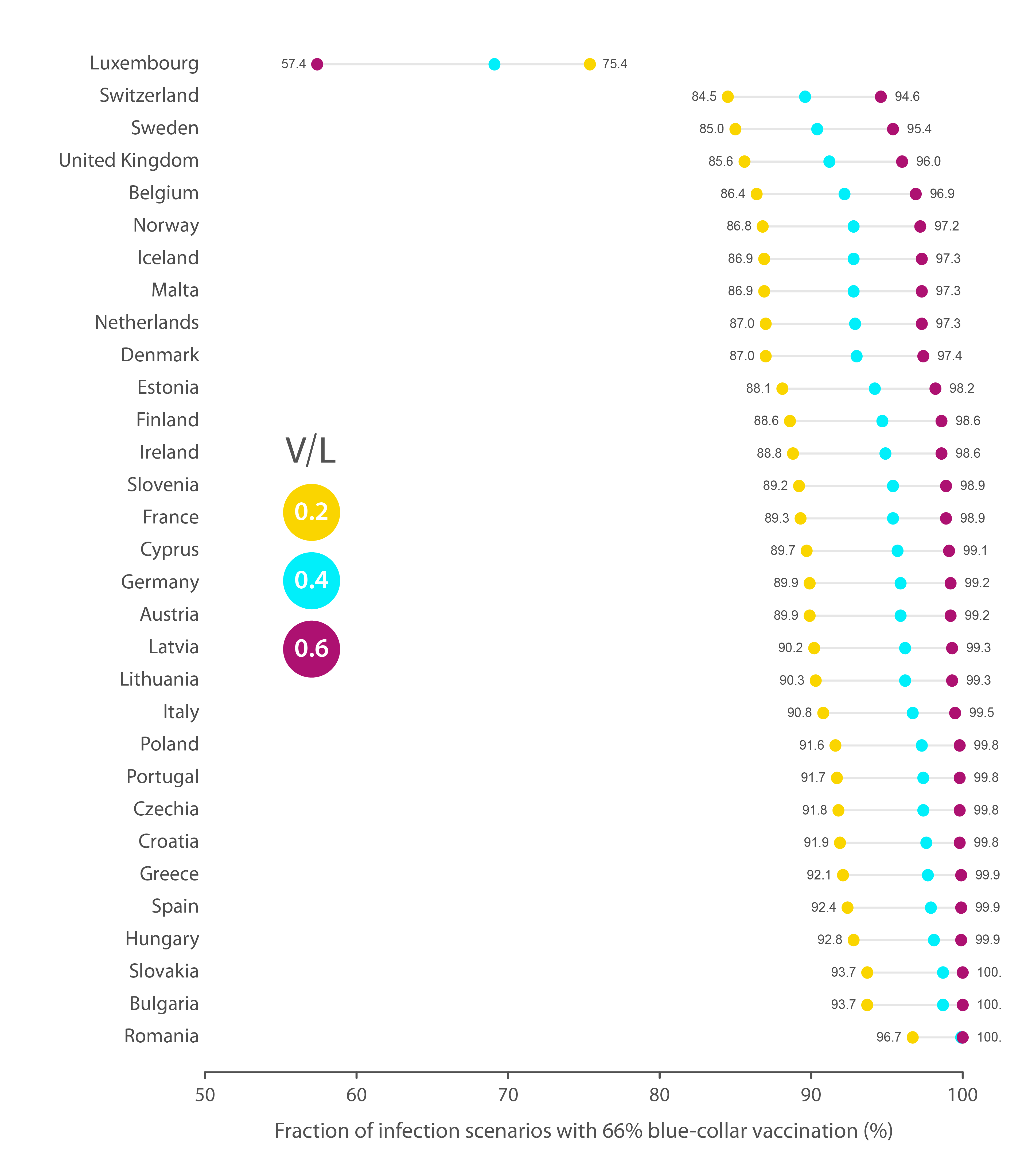}
     \caption{\textbf{The ratio of infection scenarios that require 50\% of blue-collar vaccination.} $V/L$ denotes vaccine availability for 20-40-60\% of all employees.}
     \label{fig:SI_fig1}
\end{figure} 

\end{document}